\documentclass{emulateapj}

\def\deg{$^{\circ} $}
\newcommand{\minusone}{$^{-1}$}
\newcommand{\kms}{km~s$^{-1}$}
\newcommand{\kmsm}{km~s$^{-1}$~Mpc$^{-1}$}

\newcommand{\Ha}{$\rm H\alpha$}
\newcommand{\hi}{\ion{H}{1}}
\newcommand{\nii}{\ion{N}{2}}
\newcommand{\iband}{{\em I}-band}
\newcommand{\bband}{{\em B}-band}
\newcommand{\Mi}{$M_{\rm I}$}
\newcommand{\rmax}{$R_{\rm max}$}
\newcommand{\ropt}{$R_{\rm opt}$}
\newcommand{\rd}{$r_{\rm d}$}
\newcommand{\rrd}{$r/r_{\rm d}$}
\newcommand{\rropt}{$r/R_{\rm opt}$}
\newcommand{\rpe}{$r_{\rm PE}$}

\newcommand{\about}{$\sim$}

\slugcomment{Accepted for publication in the Astrophysical Journal}

\shorttitle{Template Rotation Curves}
\shortauthors{Catinella et al.}

\begin{document}

\title{Template Rotation Curves for Disk Galaxies}
\author{Barbara Catinella\altaffilmark{1}, Riccardo Giovanelli\altaffilmark{2}, 
\& Martha P. Haynes\altaffilmark{2}}

\altaffiltext{1}{National Astronomy and Ionosphere Center, Arecibo Observatory, 
HC3 Box 53995, Arecibo, PR 00612, USA; bcatinel@naic.edu.
The National Astronomy and Ionosphere Center is operated by Cornell University 
under a cooperative agreement with the National Science Foundation.}
\altaffiltext{2}{Center for Radiophysics and Space Research and National Astronomy and 
Ionosphere Center, Cornell University, Ithaca, NY 14853, USA;
riccardo@astro.cornell.edu, haynes@astro.cornell.edu.}

\begin{abstract}
A homogeneous sample of $\sim$2200 low redshift disk
galaxies with both high sensitivity long-slit optical spectroscopy and
detailed \iband\ photometry is used to construct average, or template,
rotation curves in separate luminosity classes, spanning 6 magnitudes
in \iband\ luminosity. The template rotation curves are expressed as
functions both of exponential disk scale lengths \rd\ and of
optical radii \ropt, and extend out to 4.5--6.5 \rd, depending on
the luminosity bin.
The two parameterizations yield slightly different results beyond
\ropt\ because galaxies whose \Ha\ emission can be
traced to larger extents in the disks are typically of higher
optical surface brightness and are characterized by larger values
of \ropt/\rd. By either parameterization, these template rotation curves
show no convincing evidence of velocity decline within the spatial
scales over which they are sampled, even in the case of the most luminous
systems. In contrast to some previous expectations, the fastest rotators
(most luminous galaxies) have, on average, rotation curves that are flat
or mildly rising beyond the optical radius, implying that the dark
matter halo makes an important contribution to the kinematics
also in these systems.
The template rotation curves and the derived functional fits provide
quantitative constraints for studies of the structure and evolution of
disk galaxies, which aim at reproducing the internal kinematics
properties of disks at the present cosmological epoch.

\end{abstract}


\section{Introduction}\label{s_intr}

The determination of the rotational characteristics of galaxy disks is
fundamental to understanding the role that dynamics plays in galaxy
formation and evolution over cosmic time. Over the last decade, a
large number of studies have addressed the issue of disk formation and
evolution in current hierarchical cosmologies by means of analytical or
semi-analytical theoretical models \citep[e.g.,][]{dss97,mo98,mao98,mm04,afh98,fa00,vdb01,vdb02}
and numerical smooth particle hydrodynamic simulations \citep[e.g.,][]{vdb02b,ab03,gov04}. Other
groups have investigated more specifically the characteristics of the
density profiles of dark matter halos, which directly determine the
observable rotation curves (RCs) of present-day disks, using $N$-body
simulations \citep[e.g.,][]{nfw96,nfw97,bul01a,bul01b}.
A detailed characterization of the kinematic properties
of observed disk galaxies at low redshift is of key importance to
validate such models and place constraints on the properties of dark
matter halos. In this context, special attention has been recently
devoted to dwarf and low surface brightness galaxies, i.e. systems
with an internal kinematics that is thought to be dominated
by dark matter even at small radii, thus allowing to best probe the
inner structure of their parent halos (e.g., \citealt{sgh05} and references therein).
For high surface brightness galaxies, the kinematics is easier to
measure, but the presence of a more prominent stellar disk complicates
the analysis of the RC, whose decomposition into a dark matter and a
disk component is generally not unique 
\citep[see, for instance,][and references therein]{jvo03,dcd05}.

The derivation of {\em average} RCs representative of the kinematic properties
of a large number of local spirals is also important on its own,
to provide a standard reference against which similar, distant samples can be 
compared. Furthermore, statistical descriptors of the amplitude and
radial variation of rotation velocity within disks would benefit all
the applications that rely on the assumption of a RC model, such as 
determinations of rotational widths from spectroscopic data that lack 
velocity or spatial resolution. Examples of the latter include
internal kinematic studies of intermediate redshift spirals through spatially 
resolved optical spectroscopy, where synthetic RCs are needed 
in order to estimate the full velocity width
\citep[e.g.,][]{vog97,sim99,zie03,boh04}.

Beginning perhaps with \citet[]{rr73}, many previous works
have explored the variation in rotation curve form among galaxies. As
larger samples of rotation curves have been amassed, attempts have been made to 
produce more quantitative descriptions of
empirically--derived average, or {\em template} RCs.
Notably, average RCs binned by luminosity and with radial distances expressed
in units of the optical radius \ropt\ defined as the radius encompassing
83\% of the total integrated light were previously presented 
by \citet[hereafter PSS96]{pss96}
for a sample of 616 spiral galaxies. This definition of \ropt\
is equivalent to 3.2 \rd\ for a pure exponential disk.
PSS96 showed that their average RCs
were well described by an analytical form, obtained as the combination
in quadrature of a dark matter halo and an exponential disk, with fit
coefficients depending on galaxy luminosity only. Based on these
results they claimed the existence of a universal relation, the 
{\em universal rotation curve} (URC), whereby the shape and amplitude
of an {\em observed} RC at any radius is completely determined by the
galaxy luminosity, as originally proposed by \citet{ps91}. 
According to the URC predictions, the most luminous systems have RCs
that peak at $\sim$0.8 \ropt\ and decline beyond that radius, implying
a minor contribution of the dark matter halos to the internal
kinematics of those galaxies. It should be also noticed, however, that
PSS96's average RC tracings beyond \ropt\ are based on linear
extrapolations of outer velocity gradients as functions of luminosity
mostly obtained from \hi\ data.
Several studies have discussed the inadequacy of the URC parameterization
\citep[e.g.,][]{cou97,ver97,wil99,gma04}. 

The purpose of this paper is to provide a set of template RCs in bins 
covering a wide range of galaxy luminosity and based on an extensive set of high
sensitivity long-slit optical RCs, for which a homogeneous body of
high quality \iband\ photometry is also available. Our intent is not
to propose a new universal relation, but rather to obtain a more reliable
characterization of the average RCs of late type, high brightness disk
galaxies by taking advantage of a much larger data set, thereby negating the
need for extrapolation within the optical disk.
In order to allow a direct comparison 
with PSS96's average RCs and URC models, we provide template RCs 
parameterized as functions of \ropt. Furthermore, we also present a
set of templates which
parameterize the distance from the center of a galaxy by means of 
the exponential disk scale length \rd, a quantity naturally adopted in
models of galaxy disks.

This work makes use of a large
spiral galaxy photometric and spectroscopic dataset, dubbed SFI++
(see \S~\ref{s_sample}), compiled
to investigate the internal kinematics of disk galaxies at low
redshifts, as well as to study the peculiar velocity field in the
local universe via application of the Tully-Fisher
\citep[TF;][]{tf77} method.
Details on the extraction of related quantities from long-slit
spectra, including our RC folding technique and algorithm for the
measurement of rotational velocities for TF applications,
can be found in \citet[hereafter CHG05]{chg05}.

This paper is organized as follows. In \S~\ref{s_sample} we describe
our data sample and illustrate the procedure used to derive template
RCs. The results obtained using the parameterizations in terms of
exponential disk scale lengths or optical radii are presented and
compared in \S~\ref{s_res}. In order to explain systematic differences
between the outer slopes of the two sets of template RCs, the
correlation between \ropt/\rd\ ratio and RC extent observed for the
galaxies in our sample is also investigated in some detail. The impact
of internal extinction on the inner slopes of RCs is analyzed in
\S~\ref{s_ext}, where template RCs are computed for separate intervals
of galaxy inclination. The comparison with previously published
results, most notably the URC models, is discussed in \S~\ref{s_urc},
and our conclusions summarized in \S~\ref{s_concl}.
A value for the Hubble constant of $\rm H_0=70$ \kmsm\ is assumed 
throughout this work.

\section{Data Sample and Template RC Derivation}\label{s_sample}

A compilation of \iband\ photometry and rotational width data (obtained
from long-slit optical rotation curves and/or spatially integrated
\hi-line profiles) for several thousand spiral galaxies has been assembled by our 
group with the principal purpose of mapping the peculiar velocity
field in the local Universe. This sample, referred to as SFI++,
includes several already published data sets \citep[CHG05;][and
references therein]{shg05,vog04,dal00} along with recently obtained
observations; details of this compilation will be fully discussed
elsewhere \citep[]{klm05,cms05}.
Most notably, SFI++ includes the southern spiral galaxy
sample of Mathewson and collaborators \citep[MFB sample;][]{mfb92,mf96}. 
The \iband\ isophotal photometric fits and RCs derived from long-slit
optical spectroscopy for the MFB sample are available to us in digital 
form, and were reprocessed to achieve consistency with the rest of
SFI++ (photometric quantities and rotational widths have been
extracted according to the methods described in \citealt{hay99} and
CHG05, respectively).

In this work, we use the subset of galaxies in the SFI++ sample for
which high quality long-slit \Ha/[\nii] RCs and \iband\ photometry 
are available. Long-slit spectra are of high quality when an extended
RC can be reliably extracted, i.e. when the \Ha\ emission line has
high signal-to-noise, is extended, and allows to adequately trace the
shape of the RC across its spatial extent. Analogously, \iband\ images
are of high quality when photometric parameters can be reliably
extracted (i.e., photometric quality, no stars superimposed on the
galaxy, object not close to the edge of the frame, etc. Also discarded
are cases in which the images are of high quality  but the fits are
poor because of the presence of a bar, asymmetries, or other
problems). Template RCs were obtained by binning the galaxies in
\iband\ luminosity intervals, and computing the average RC in each
bin as described below. Two separate sets of solutions were
derived by adopting different spatial parameterizations, in which the
radial coordinates of the RCs are expressed in units of the \iband\
disk scale length, \rd, or the optical radius, \ropt\ (i.e., the
radius within which 83\% of the total \iband\ light is included, as in
PSS96. Notice that \ropt\ is a cumulative quantity and not an
isophotal radius). The template RCs were then fitted with the analytic
function:

\begin{equation}
        V_{\rm PE}(r) = V_0(1-e^{-r/r_{\rm PE}})(1+\alpha r/r_{\rm PE})
\label{eq_polyex}
\end{equation}
\noindent
referred to as the {\em Polyex} model \citep[]{gh02}, where $V_0$, \rpe, 
and $\alpha$ respectively determine the amplitude, the exponential 
scale of the inner region, and the slope of the outer part of the RC
(for template RC fitting, $r$ and \rpe\ are expressed in units of \rd\
or \ropt).
Equation~\ref{eq_polyex} is an empirical expression that fits well
a large variety of RC shapes (including those declining at large
radii), as discussed in CHG05.
Alternative fitting functions have been proposed by other authors
\citep[e.g.,][]{cou97,kkbp98,vog04}; since they are all empirical, trade-offs between
simplicity and detail result. A comparison between Polyex and other
fitting expressions is beyond the scope of this paper and will not be
discussed here.

All the individual RCs used for the derivation of the template curves
were fitted using Equation~\ref{eq_polyex} and, although their Polyex
parameters were not used in this work, a quality code based on the
goodness of the fit was used to select those included in the final
sample. For reasons that will be clarified later,
the following objects were also not used to construct
template RCs: (a) galaxies with inclinations to the line of sight
$i\geq80^\circ$ (see also \S \ref{s_ext}); (b) galaxies with
absolute \iband\ magnitude lying outside the [$-$24.0,$-$18.0]
interval; (c) RCs with spatial extent smaller than 2 \rd\ or 
0.6 \ropt, for the \rd\ or \ropt\ parameterizations, respectively; 
and (d) RCs with average velocity (computed over the spatial interval
$\Delta r$, as defined in [3] below) incompatible with the value
obtained using all the RCs in the same luminosity bin (over the same
$\Delta r$). Most of the rejected objects fell into category [a] (472
out of 2777 galaxies in the initial sample). Also, because of the
selection criterion [c], the sizes of the final data sets used to
derive \rd\ and \ropt\ template RCs, hereafter referred to as 
``\rd\ and \ropt\ samples'', were slightly different, including 2155
and 2169 galaxies respectively.\\

After binning the galaxies by luminosity, template RCs were obtained
as follows: \\
(1) {\em Folding} -- Each RC was folded about its center of symmetry,
    as determined by fitting a URC model (PSS96) to the unfolded
    RC. As pointed out elsewhere (see references in \S~\ref{s_intr}), the URC
    parameterization is inadequate to model individual RCs, but it
    approximates the overall RC shape well enough for the purpose of
    determining its center of symmetry, as argued in CHG05, where our
    folding technique is described in detail. \\
(2) {\em Rescaling and Resampling} -- The radial variable of each
    folded RC was scaled by the \rd\ (\ropt) value of the galaxy, and
    the RC amplitude was deprojected to the edge-on view. Inclinations
    were obtained from ellipticities measured from isophotal fits to
    the \iband\ images, and corrected for seeing effects, following the
    prescriptions of \citet{gio97a}.
    RC amplitudes were then resampled on a fixed spatial grid of spacing
    0.1 \rd\ (0.03 \ropt) using a linear interpolation including the
    four observed points nearest to the grid position. This assures
    that RCs of nearer galaxies, with denser spatial sampling, do not
    bias the average RCs. We have tested that the results are not
    sensitive to the choice of the spatial bin centers. Only the
    resampled RCs were used after this point. \\
(3) {\em Normalization} -- Obtaining an average RC for each luminosity
    class requires attention to the fact that the RC of each galaxy is
    sampled out to a different maximum radius \rmax. To determine the
    average amplitude of the template RC of a given luminosity class,
    we thus used only the section $\Delta r$ between 2\rrd\ and
    3\rrd\ (0.6--0.9 \rropt) of each RC. Over this interval, nearly
    all RCs are well sampled (91\% of the RCs in the \rd\ sample
    and 95\% of those in the \ropt\ one extend beyond 2.5 \rrd\
    and 0.75 \rropt, respectively) and exhibit relatively mild
    gradients. Using a range of radii rather than a single radial
    position also limits shot noise effects. The normalization of the
    RCs less extended than 3 \rd\ (0.9 \ropt) was performed using
    the available points within $\Delta r$ (29 RCs in the \rd\ sample
    were normalized using the velocity at 2.0 \rd). For each RC, we
    thus measured the average rotational velocity $v_{\Delta r,i}$;
    objects without a meaningful determination of that parameter have
    been excluded from our final sample. For a given luminosity class
    $M$, the template RC was obtained by scaling each RC by the ratio
    $R_{v,i}\equiv v_{\Delta r,i}/\langle v_{\Delta r,M}\rangle$, where 
    $\langle v_{\Delta r,M}\rangle$ is the average $v_{\Delta r,i}$
    for the class (objects with $R_{v,i}\geq 2$, corresponding to a
    5$\sigma$ deviation from the mean of the distribution for the
    whole sample, were discarded). The 1$\sigma$ dispersion of the
    $v_{\Delta r,i}$ velocities is, on average, 17\% of 
    $\langle v_{\Delta r,M}\rangle$ (it varies between 14\% and 22\%
    among the different luminosity classes; $\langle v_{\Delta r,M}\rangle$ 
    values are listed in Tables~\ref{mbins_rd} and \ref{mbins_ropt},
    presented in \S \ref{s_res}).
    We refer to the RCs scaled in this way as {\em normalized} RCs.\\
(4) {\em Averaging} -- Mean velocities and variances were computed
    (after clipping at the 2$\sigma$-level) for each grid value of
    \rrd\ (\rropt) for which there were at least 4 valid samples, and
    for each luminosity class. The choice of 4
    samples represents a good compromise between maximizing the
    number of radial bins for the computation of the template RCs
    and keeping enough points within each radial bin to calculate a
    meaningful average velocity. Since the
    scatter of the RCs around the template is artificially reduced
    over and near the interval $\Delta r$ where the velocity
    normalization was performed, we replaced the velocity
    variance at each grid point (over the entire spatial range covered
    by the template RC, not just $\Delta r$) with that computed using
    the unnormalized RCs. In other words, the template RC points were
    computed using normalized RCs, but the corresponding
    dispersions are those of the {\em unnormalized} RCs.\\
(5) {\em Smoothing} -- The resulting average RCs were
    Hanning-smoothed, and every other point was recorded,
    yielding a grid spacing of 0.2 \rd\ (0.06 \ropt).
    As mentioned above, each template RC point was computed by averaging at
    least 4 samples (mean velocities resulting from the
    average of 3 or less points were not recorded).
    The final template RC points were obtained by expanding
    underpopulated spatial bins (i.e., bins resulting from the average of
    $4 \leq N <10$ individual RC points) in order to include at least 10 samples.
    Positions and error bars of
    the expanded bins are given by the average of the corresponding
    quantities for the replaced bins.

The binning into luminosity classes was initially done with a constant
step $\Delta M$=0.4. The bins at the faint end of the
interval were expanded, however, to avoid the overlap of the
corresponding template RC solutions; in particular, two classes 
$M_{\rm i}$ and $M_{\rm j}$ were combined if 
$\vert \langle v_{\Delta r,M_{\rm i}}\rangle - \langle v_{\Delta
  r,M_{\rm j}}\rangle \vert \leq 10$ \kms.
The final 10 luminosity classes span the \iband\ absolute magnitude
interval [$-$24.0,$-$18.0] (18 objects with luminosities outside this
range were discarded).

The derivation of the template RCs was identical for the \rd\ and \ropt\
parameterizations, except for the values of the grid spacing and
the normalization interval $\Delta r$ (as in [2] and [3] above).
For the \ropt\ template RCs, grid spacing and $\Delta r$ were chosen
to match the corresponding \rd\ ones, taking into account that the
average value of the \ropt/\rd\ ratio is 3.31$\pm$0.01 for the sample
used to construct \rd\ template RCs.

As discussed in \S~\ref{s_ext}, the inner slope of observed \Ha/[\nii]
RCs depends on inclination, due to internal extinction in the inner
regions of disks. The impact of dust attenuation is more severe for edge-on
disks and for intrinsically brighter systems. For this reason, 472 galaxies
with inclination $i\geq80^\circ$ were excluded from the analysis, as
already mentioned.
Seeing may also systematically bias the inner slopes of 
observed RCs, especially for more distant galaxies. The set of
galaxies used to construct the template RCs consists of relatively
nearby, large spirals, of median recessional velocity
$\sim$7000 \kms\ and isophotal radius $r_{I23.5} \sim 30$\arcsec.
We investigated a possible dependence of the shape of RCs on
galaxy distance, and found no significant effect for objects in
our sample. As another way of testing the impact of seeing on RC
shape, we divided our sample into 3 intervals of (linear) disk scale
length $R_{\rm d}$ ($R_{\rm d} < 2$ kpc, 2 kpc $\leq R_{\rm d} < 3$
kpc, and $R_{\rm d} \geq 3$ kpc) and recomputed the template RCs
forcing the same velocity normalization obtained for the whole \rd\
sample (as done in \S~\ref{s_ext} for the inclination classes). We
found that galaxies with larger $R_{\rm d}$'s tend to have slightly
larger rotational velocities than ones with smaller $R_{\rm d}$'s;
however the effect is small (well within the 1$\sigma$ error bars on
the template RC velocities. In
particular, the maximum velocity difference is of the
order of 20 \kms\ or less for the template RCs in the central
luminosity bins, $-22.8 < M_{\rm I} < -20.8$; there are no galaxies
with $R_{\rm d} < 2$ kpc in the two brightest bins, nor ones with
$R_{\rm d} \geq 3$ kpc in the two faintest bins. Larger velocity
differences are observed for the $M_{\rm I}= -22.8$ case, but the
$R_{\rm d} < 2$ kpc result is uncertain due to small number
statistics) and limited to the $r<$\rd\ spatial interval.

\section{Template RCs: Results and Comparison of The Two Parameterizations}\label{s_res}

\begin{figure}
\plotone{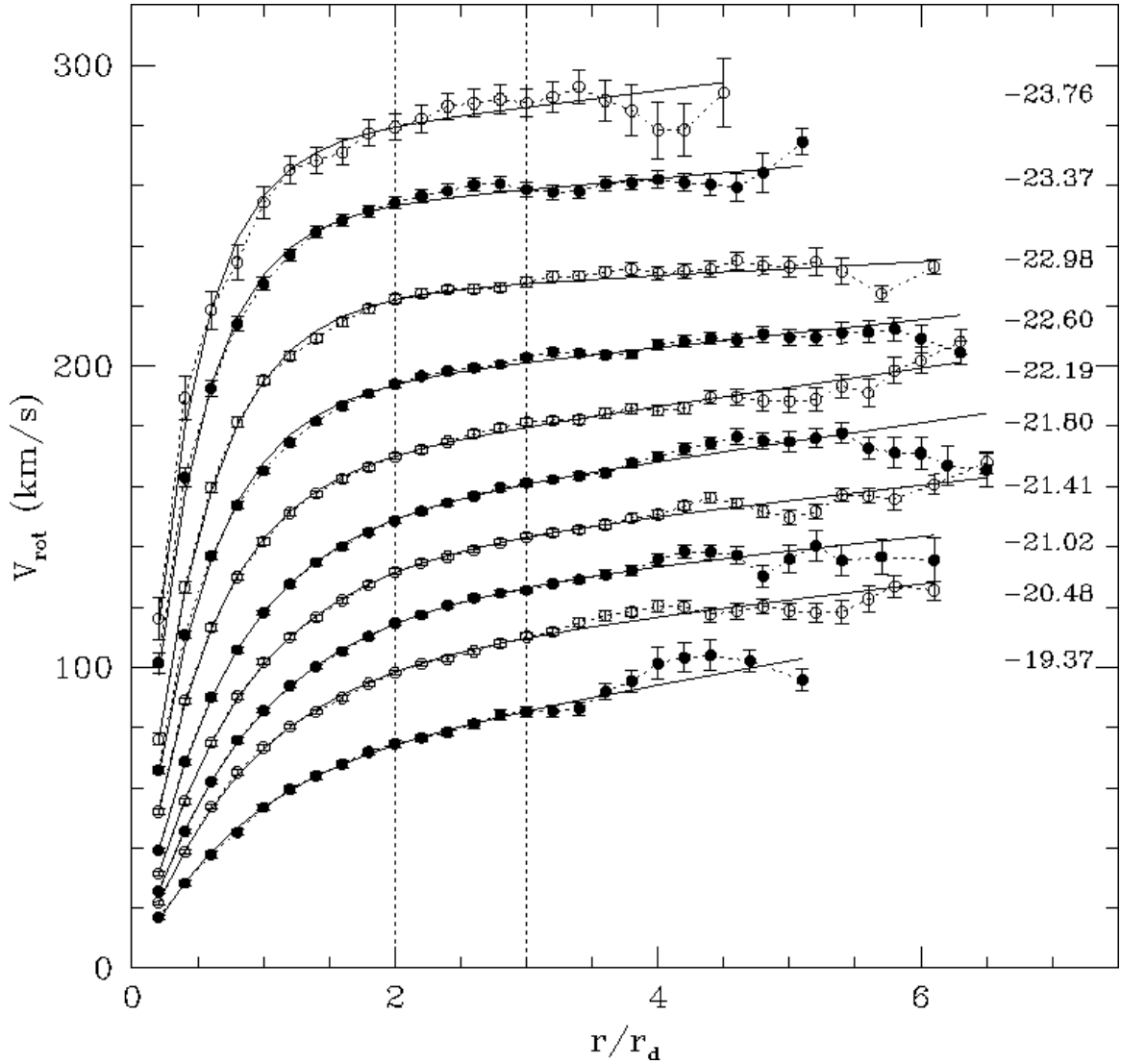}
\caption{Template RCs parameterized as functions of exponential disk
  scale lengths. Each curve is labeled on the right by its mean \iband\
  absolute magnitude ($\rm H_0=70$ \kms\ Mpc\minusone). The final
  sample includes 2155 RCs extending beyond 2 \rd, and with
  inclination to the line of sight $i<$80\deg. The vertical, dotted
  lines show the interval over which the velocity normalization was
  performed (see \S \ref{s_sample}). The error bars are Poissonian
  errors on the mean. Polyex fits to the data points
  are indicated by solid lines; the fit coefficients are presented in
  Figure~\ref{pe_parms_rd} and Table~\ref{mbins_rd}.
\label{avgrcs_rd}}
\end{figure}

Figure~\ref{avgrcs_rd} shows the template RCs expressed as functions of
exponential disk scale lengths \rd, obtained by averaging the data in
each of 10 luminosity intervals. Each curve is labeled by its mean
\iband\ absolute magnitude (as listed in column 3 of
Table~\ref{mbins_rd}, described below); the interval 
$\Delta r$ over which the velocities are normalized 
within each luminosity class is indicated by dotted lines.
For each template, the outermost spatial bins have been expanded to
include at least 10 elements; hence their spacing from adjacent points is not
exactly 0.2 \rrd. The error bars are Poisson errors on the mean,
calculated using unnormalized RCs (see [4] in \S~\ref{s_sample}).
Notice that the error bars on the outermost points may not be
accurate due to small number statistics.
Polyex (see eq.~\ref{eq_polyex}) fits to the template RC data points are
shown as solid lines. Our results show a smooth transition of RC shapes
from small to more massive (luminous) systems, with the latter being
characterized by a steeper initial velocity rise and a flatter outer
slope. Even in the case of the most luminous bins we see no convincing evidence of
velocity decline {\em within the spatial scales sampled here},
contrarily to what expected based on the URC predictions. In fact, while
the very outermost points of the template RCs show a small velocity
decrease in a few cases (e.g., those labeled $-$19.37 and $-$22.60 in
the figure), there is no systematic trend that indicates a falling RC
within at least 5 disk scale lengths (6 for the central luminosity
bins). In other words, if we were to determine the outer slope of each
template RC by fitting a first order polynomial beyond 2 or 3 \rd,
none of the fits would have a negative slope.
Notice also that the curve labeled $-$21.80 appears to decline beyond
\about 5 \rd, but the two adjacent templates ($-$21.41 and $-$22.19)
have rising outer slopes.
This issue will be further discussed in \S~\ref{s_urc}, where
we compare our average RCs with those derived by PSS96, upon
which the URC parameterization is built. 
Lastly, we point out that the {\em bump} at large radii seen for the
lowest luminosity template RC is caused by a lack of faint galaxies
with extended RCs. In fact, bright galaxies in that luminosity bin
(i.e., those with $M_{\rm I} < -19.5$) span all values of RC extent
(up to 6 \rd), whereas fainter ones ($M_{\rm I} > -19$) have RCs that
extend up to 4 \rd\ at most (with only one exception, a RC with 
$M_{\rm I} = -18.6$ and \rmax = 4.8 \rd).

\begin{figure}
\plotone{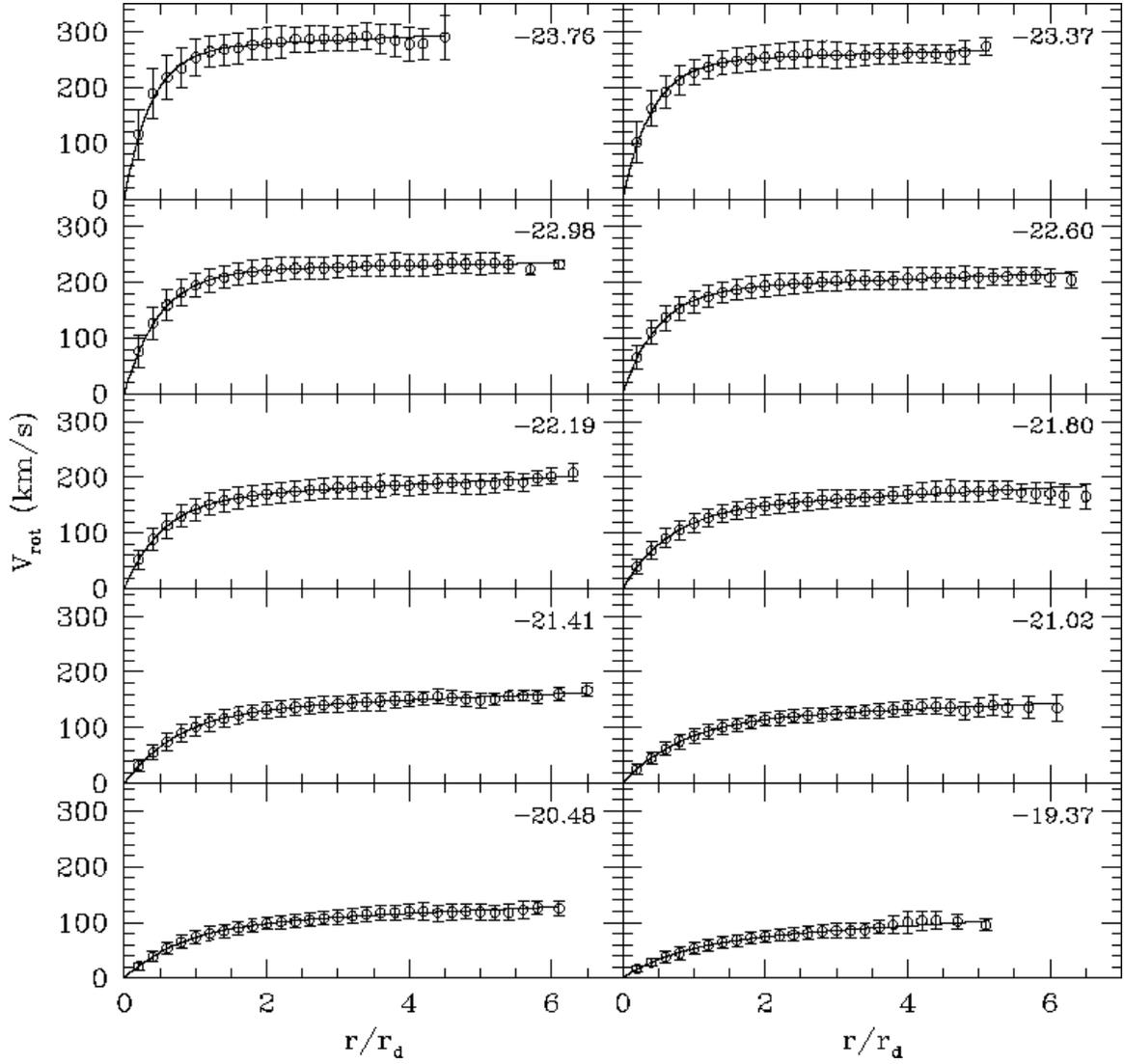}
\caption{The template RCs shown in Figure~\ref{avgrcs_rd} are
  displayed here in separate panels, with error bars representing the
  dispersion around the mean. The solid lines indicate the Polyex fits
  from Figure~\ref{avgrcs_rd}.
\label{avgrcs_rd_stdv}}
\end{figure}

In order to illustrate the actual spread in RC shapes and amplitudes
within each luminosity class, we include a version of Figure~\ref{avgrcs_rd} in
which the error bars represent the dispersion around the mean
(Figure~\ref{avgrcs_rd_stdv}). Typical values of the standard
deviation around the mean are 13 \kms\ for the template RCs in the
lowest 4 luminosity bins, 30 \kms\ for the one labeled $-$23.76, and
20 \kms\ for the others.

Table~\ref{mbins_rd} summarizes the statistics for each luminosity
class and provides the values of the Polyex parameters of the template RC
fits shown in Figures~\ref{avgrcs_rd} and \ref{avgrcs_rd_stdv}. The
quantities listed are: \\
{\em Column (1):} center of the luminosity bin, \Mi, in magnitude units.\\
{\em Column (2):} width of the luminosity bin, $\Delta M_{\rm I}$.\\
{\em Column (3):} mean value of the absolute \iband\ magnitude within
  the corresponding bin, $\langle M_{\rm I} \rangle$.\\
{\em Column (4):} mean velocity of the RCs, $\langle v_{\rm \Delta r,M} \rangle$, 
  measured between 2\rd\ and 3\rd. This value is used to normalize the
  velocities of the individual RCs belonging to the same luminosity
  bin, as explained in \S~\ref{s_sample} (item [3]).\\
{\em Column (5):} number of RCs included in each luminosity bin, $N_{\rm \Delta r,M}$. \\
{\em Columns (6)--(8):} coefficients of the Polyex model (see eq.~\ref{eq_polyex}) 
  fits to the template RCs, with errors; $V_0$ is measured in \kms,
  \rpe\ is expressed in units of \rd, and $\alpha$ is dimensionless. \\
The bin centered at \Mi$=-$23.8 contains only 43 objects, but is 
particularly interesting because the most luminous galaxies are
expected to show the most declining RCs within a few disk scale lengths
(based on URC predictions), implying a rather small contribution 
of the dark matter component to the internal kinematics of such systems. 
The fastest rotators represent also important tools to test our
understanding of galaxy formation and disk stability 
(e.g., \citealt{spe05}; see also \S~\ref{s_urc}).

\begin{figure}
\plotone{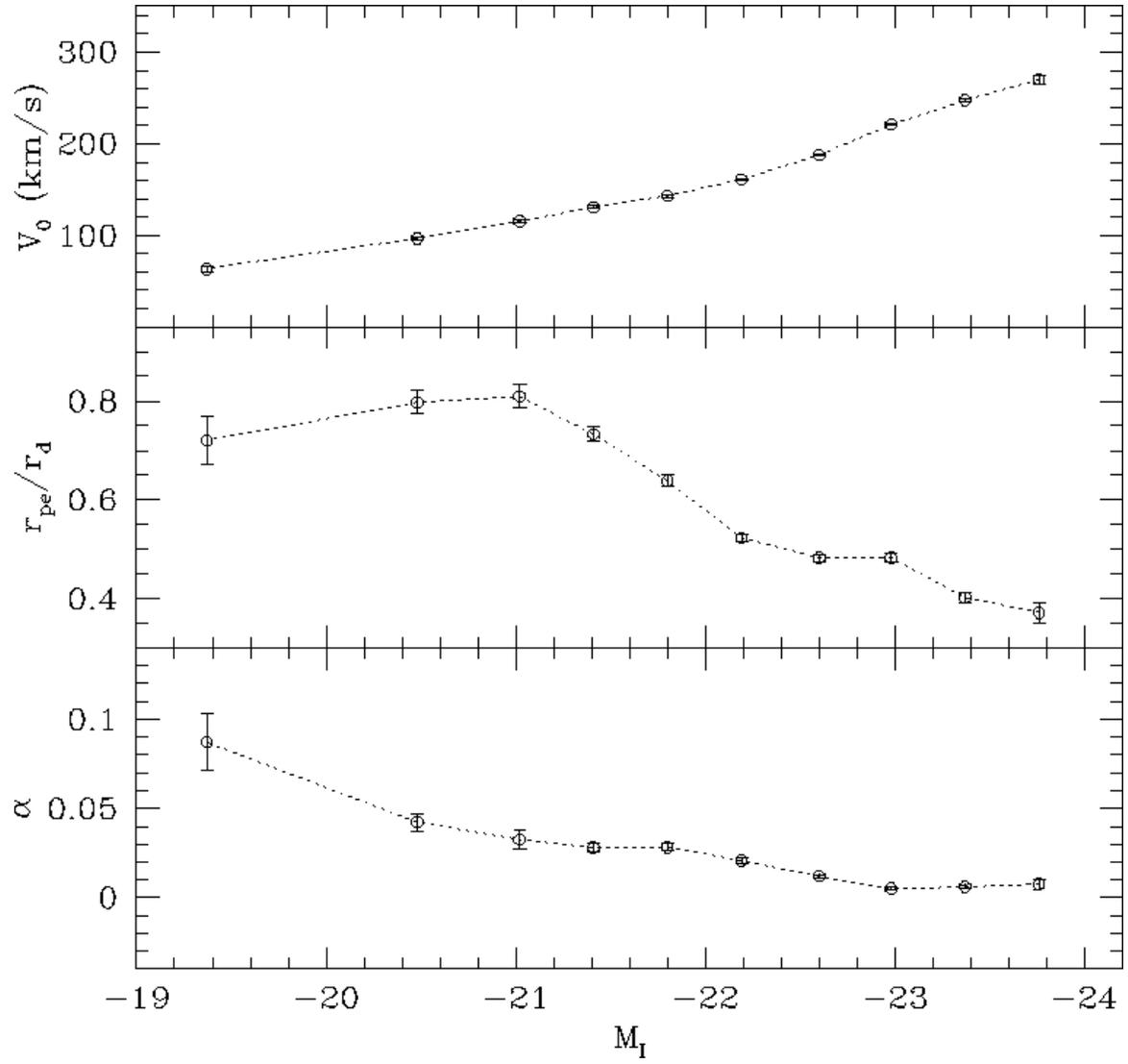}
\caption{Polyex model coefficients of the fits shown in Figure~\ref{avgrcs_rd}, 
  plotted as functions of \iband\ absolute magnitude. See also Table~\ref{mbins_rd}.
\label{pe_parms_rd}}
\end{figure}

The variations of the Polyex parameters with \iband\ absolute magnitude
are displayed in Figure \ref{pe_parms_rd}. The trend seen in the top panel
reflects the well-known result that brighter disk galaxies rotate
faster (the TF relation); the other two
panels show that RCs of brighter systems have steeper velocity
increase within the inner regions and flatter outer slopes.

The results for the template RCs expressed as functions of optical
radii rather than disk scale lengths are presented in Figures
\ref{avgrcs_ropt}, \ref{avgrcs_ropt_stdv}, \ref{pe_parms_ropt}, and
Table~\ref{mbins_ropt} (analogous to Figures \ref{avgrcs_rd},
\ref{avgrcs_rd_stdv}, \ref{pe_parms_rd}, and Table~\ref{mbins_rd} respectively). 
As mentioned in \S~\ref{s_sample}, the interval $\Delta r$=0.6--0.9
\rropt\ for velocity normalization within each luminosity class
corresponds to $\Delta r$=2--3 \rrd\ for an average \ropt/\rd\ ratio
of 3.3, as determined for the 2155 galaxies of the \rd\ sample.
The difference between the corresponding values of the
velocity normalization (listed in column 4 of Tables \ref{mbins_rd}
and \ref{mbins_ropt}), is smaller than 3 \kms\ for all the luminosity
bins except the \Mi$=-$23.8 one, where the discrepancy is slightly
larger (5.4 \kms). A quick inspection of the two tables also reveals
that the numbers of RCs belonging to the same luminosity bin are
similar, but not the same, for the \rd\ and \ropt\ samples. Such small
differences are caused by the exclusion of
RCs that are spatially less extended than 2 \rd\ or 0.6 \ropt\
and by the fact that the \ropt/\rd\ distribution is not
a delta function centered on 3.3. In fact, if we dropped the
restriction on RC extent, {\em or} if all the RCs had
\ropt=3.3 \rd, the composition of the two samples used for template
RC derivation and their distribution into luminosity classes would be
identical. In practice, for instance, a RC with a maximum extent \rmax = 2.1
\rd\ is part of the \rd\ sample, but will be included in the \ropt\
sample only if \ropt/\rd $\leq 3.5$; viceversa, a RC with \rmax = 1.9 \rd\
will be missing from the \rd\ set and included in the \ropt\ one if
\ropt/\rd $\leq 3.2$. A comparison between the distributions of
\ropt/\rd\ for the two sets of galaxies (not shown) confirms
that the \ropt\ sample has a slightly larger (smaller) fraction of
objects with \ropt/\rd $<3.3$ ($>3.3$) with respect to the \rd\
sample.

When scaled with \ropt\ rather than \rd, the template RCs have flatter outer slopes. 
As for the \rd\ solutions, there are a few examples of templates whose
outermost points show a small velocity decrease (Figure \ref{avgrcs_ropt}); 
however, there is no clear indication of declining RCs within the
spatial interval sampled by our data, in the sense that linear fits to
the outer regions of the template RCs (beyond 0.6 or 0.9 \ropt) would
have flat or positive slopes. The only exception is the curve labeled
$-$22.60, but again we notice that the adjacent templates have rising
or flat slopes.
The solutions obtained with the two parameterizations are very
similar, except for the tendency of the \rd\ template RCs to show
steeper outer slopes, as indicated by the larger values of the Polyex
parameter $\alpha$ in the bottom panel of Figure~\ref{pe_parms_rd} and
in column (8) of Table~\ref{mbins_rd}, compared with the corresponding
results for the \ropt\ case. This effect is clearly illustrated in
Figure~\ref{avgrcs}, where the two sets of Polyex fits displayed
in Figures \ref{avgrcs_rd} and \ref{avgrcs_ropt} are reproduced as
dashed and solid lines, respectively. Excluding the \Mi$=-$23.8 bin,
which is also characterized by the poorest statistics, the
template RCs obtained with the two parameterizations are nearly
indistinguishable within $r\sim$1 \ropt, whereas the \rd\ tracings
are systematically steeper than the \ropt\ ones beyond that radius.
If all the RCs had the same \ropt/\rd\ ratio, the results would be
identical, and in fact the \rd\ templates could be obtained directly
from the \ropt\ ones (or viceversa) by rescaling the horizontal axis. On the
other hand, a distribution of \ropt/\rd\ values should introduce
only scatter, and not a systematic change of RC slopes. In order to
explain the observed bias, \ropt/\rd\ must be dependent on RC extent,
in the sense that RCs sampled out to larger fractions of the disks
must also be characterized, on average, by larger \ropt/\rd.
Figure~\ref{roptrd} shows the correlation between \ropt/\rd\ and RC
extent, \rmax, measured in units of disk scale lengths, for the 2155
galaxies of the \rd\ data set; the dotted line is at \ropt/\rd = 3.3,
the average value for that sample, and the lack of points with
\rmax/\rd $<2$ is due to a selection effect, as mentioned above.
Our data show that for a fixed disk scale length, galaxies with more
extended RCs have, on average, larger optical radii, thus explaining
the systematic difference of slopes between the \rd\ and \ropt\
template RCs seen in Figure~\ref{avgrcs}.
We investigated the origin of the correlation between \ropt/\rd\ and
RC extent by binning the sample into four intervals with different
disk central surface brightness $\mu_0$ (measured from our \iband\
profiles and corrected for Galactic and internal extinction,
cosmological k-term, and converted to face-on perspective), as
indicated in Figure~\ref{roptrd_sb}, and by coding the objects based
on their morphological types (adopting the scheme in the RC3 catalog;
\citealt{dev91}). Since our sample is mostly composed of Sb or S... (42\%),
Sbc (14\%), and Sc (36\%) galaxies, we grouped the data points into
two classes only, Sb and earlier types (asterisks) and Sbc and later
types (open squares). Our results indicate that:
(a) not surprisingly, the \Ha\ emission is typically traced to a
larger extent in the disks of galaxies with higher surface brightness;
(b) the average value of \ropt/\rd\ increases with decreasing $\mu_0$,
as the distribution of data points with respect to the \ropt/\rd =
3.3 line in each panel illustrates;
(c) in the upper panel, where the data span the largest \rmax/\rd\
range, the \ropt/\rd\ ratio shows a clear dependence on RC extent;
such trend is also present in the central panels, although
less convincingly;
(d) there is no obvious dependence on morphological type, presence of an
identifiable bar (not shown), or \iband\ luminosity (not shown).
In conclusion, the systematic difference of slopes beyond $r\sim$\ropt\
between the two sets of template RCs in Figure~\ref{avgrcs} is
largely caused by the high surface brightness part of our sample. In
the region between 4 and 6.5 \rmax/\rd, where the discrepancy is
larger, the contribution to the template RC points is dominated by the
objects in the two upper panels of Figure~\ref{roptrd_sb}, which are
characterized by an \ropt/\rd\ ratio that is systematically larger
than the average of the sample, especially for increasing values of
\rmax/\rd. 

In this section, we have presented template RCs expressed 
separately as functions
of exponential disk scale lengths or optical radii. While the
description in terms of \rd\ is preferred for its more immediate
interpretation in the context of galaxy structure and disk
modeling, the \ropt\ parameterization is also provided in order to
allow a direct comparison between PSS96's and our results (see \S~\ref{s_urc}).
From an observational point of view, individual values of \rd\ carry
significant errors because of the difficulty of fitting exponential
slopes in the presence of both spiral structure and radially-dependent
extinction. Being a cumulative parameter, \ropt\ is more straightforward
to measure and is less subject to extinction effects. The complicating
presence of a bulge would bias both measurements, in the sense that
the estimated values would be systematically smaller than those
determined for a galaxy with the same disk but no bulge. The
impact on \ropt\ would be direct and much larger, possibly introducing
morphological and/or luminosity effects.

\section{Impact of Internal Extinction on Template RCs}\label{s_ext}

As mentioned in \S~\ref{s_sample}, galaxies with inclination $i\geq 80^\circ$ were
excluded from the sample used to derive the template RCs, in order to minimize the effects
of the internal extinction on our results. Here we address this issue in more detail, and 
motivate our choice of the inclination threshold for inclusion in the final sample.

Several works have previously presented evidence for internal extinction in the inner
regions of disk galaxies, based on the analysis of optical and near-infrared photometry or 
spectroscopy. Extinction affects an observed RC by causing its slope
to appear shallower; this effect, first noted by \citet{gr81}, is
illustrated by the radiative transfer simulations of \citet{bos92} and
\citet{baes03}. These studies show that dust attenuation has a severe
impact on the observed kinematics of edge-on disks, even for modest
optical depths, and that such effects are strongly reduced for galaxies
that are more than a few degrees from edge-on, becoming entirely negligible 
for intermediate inclinations.
These results were confirmed by the analysis of \citet[hereafter GH02]{gh02}, 
which also showed that the prominence of the effect increases with
disk luminosity. Exploiting a substantial fraction of the SFI++ sample
used in this work, GH02 showed that the kinematic scale length \rpe\ of the
inner slope of a RC (as obtained from the Polyex model in eq.
\ref{eq_polyex}) starts increasing noticeably for inclinations
$i\gtrsim 70^\circ$, the increase being larger for more luminous
disks. By contrast, the outer slope $\alpha$ of the observed RCs shows
no dependence on inclination. 
Adopting a simple model for the dust and gas distribution in spiral
disks, \citet{vg04} were able to reproduce the observed trends in GH02
fairly well. 

In order to evaluate the impact of internal extinction on our results, we divided
our sample, which includes 472 highly inclined ($i \geq$80\deg) systems,
into 3 inclination intervals and recomputed the template RCs. 
The results are displayed in Figure \ref{incl} for the
parameterization in terms of exponential disk scale lengths; only the inner
regions are shown. Each panel corresponds to one of the \iband\
luminosity classes listed in Table \ref{mbins_rd}, from \Mi=$-$23.8
(a) to \Mi=$-$19.0 (j); different inclination bins are indicated as
solid (30\deg $\leq i <$70\deg), dotted (70\deg $\leq i <$80\deg), and 
dashed (80\deg $\leq i \leq$90\deg) lines.
The velocity normalization of the individual RCs in each luminosity
class was forced to be the same as that used for the whole \rd\
sample, as listed in column 4 of Table \ref{mbins_rd}.
Figure \ref{incl} clearly shows that: (1) the inner slope of the most edge-on
systems is systematically shallower, and (2) the effect is more pronounced
for more luminous galaxies and almost disappears for intrinsically
fainter systems.
In the case of highly inclined galaxies, additional geometric effects
can cause the inner slopes of observed RCs to appear shallower. Such
effects depend on the thickness of the disk (higher values of disk
thickness result in shallower RCs), which is a poorly known quantity,
and can account for part of the difference between the highest
inclination bin and the lower inclination ones in Figure \ref{incl}. 
The interplay between optical depth at the disk center and disk
thickness is illustrated in \citet{vg04}.

Our results confirm the previous findings of GH02 (as expected, since
they provide a different representation of the same effect for
essentially the same data set) and of the other studies mentioned
above, and show the importance of excluding the highly inclined disks
from the derivation of the template RCs.
In particular, our chosen threshold $i= 80^\circ$ allows us to
effectively eliminate the RCs that are most affected by dust
attenuation (and geometric effects related to the thickness of the
disk), without discarding an excessive fraction of the sample. \\

\section{Comparison with Previous Results}\label{s_urc}

A similar derivation of average RCs of disk galaxies binned in
\iband\ luminosity intervals was undertaken by PSS96. Their {\em sample B}
included 616 RCs extending out to $r\geq$0.8~\ropt\ (about 200 of which
beyond \ropt), obtained by \citet{ps95} by deprojecting, folding, and
smoothing the \Ha\ data of \citet{mfb92}. The RCs in this sample were
normalized to the velocity measured at the radius encompassing 65\% of
the integrated light (approximately corresponding to 2.2 \rd\ or 0.7
\ropt\ for a pure exponential disk), separated into 11 luminosity
intervals, and averaged in radial bins of size 0.1 \ropt. 
In order to characterize the velocity field of galactic disks beyond
the optical radius, this data set was complemented by {\em sample A},
a compilation of 131 RCs obtained from published optical or 21 cm \hi-line
spectroscopy. Outer gradients of RCs, defined as the fractional
variation $\delta$ of the rotational velocity between 1 and 2 \ropt,
were calculated for 27 RCs in {\em sample A} extending to about 2
\ropt\ (mostly \hi\ data), and for 5 synthetic RCs obtained from the RCs
in {\em sample B}, ``suitably co-added in order to probe the
kinematics of the outermost regions.''
The variation of $\delta$ as a function of \bband\ absolute magnitude was
then used to {\em extrapolate} the average RCs out to 2 \ropt.
The resulting RC tracings were fitted with a parametric form, obtained
as the combination in quadrature of an exponential disk and a dark
matter halo, and the fit coefficients expressed as functions of
luminosity. Based on this parameterization, shapes and amplitudes of
{\em observed} RCs at any radius are completely determined by
galaxy luminosity, implying the existence of a universal RC, as
originally proposed by \citet{ps91}.

The inadequacy of the URC to model individual RCs has been 
pointed out by several authors, as mentioned in \S~\ref{s_intr}.
Here, we want to compare PSS96's average RCs in luminosity bins (see
their Figure 4) with our template RCs parameterized as functions of
optical radii. For this purpose, the Polyex fits shown in
Figure~\ref{avgrcs_ropt} are reproduced in
Figure~\ref{urc} as solid lines, overlaid on the URC data points
(kindly provided by Paolo Salucci). Since there seems to be a
systematic offset between PSS96 and our \iband\ luminosity scales
(the \Mi\ bins for the URC tracings are, from top to bottom:
$-$23.2, $-$22.3, $-$22.0, $-$21.6, $-$20.9, $-$20.5, $-$20.0,
$-$19.4, $-$18.5), but the rotational velocity range sampled by our
two samples is the same, we optimized the velocity match at
the optical radius by shifting all the URC data points by $+$10 \kms.
As mentioned in \S~\ref{s_sample}, \citet{mfb92} data set is included
in our sample, but was reprocessed by us in the same fashion as the
rest of SFI++. Therefore the discrepancy between absolute magnitudes
might be due to a number of factors, including apparent magnitude
measurements, extinction corrections, and distance estimates to the
objects (notice that the value of the Hubble constant adopted by
PSS96, 75 \kmsm, is very similar to the one used in this work).
The comparison between PSS96 and our average RCs should be carried out
separately for the two spatial intervals below and above \ropt, due to
the different nature of the corresponding URC points, as summarized at
the beginning of this section. Below \ropt, where the URC points were
obtained by averaging RC data, the agreement is fairly good,
especially for the 5 lower luminosity bins (in the case of the curves
labeled $-$22.2 and $-$22.6, one should interpolate between the two
nearest template RCs in order to obtain one that matches the URC
velocity at \ropt). There is however a systematic tendency for the URC
tracings to have shallower slopes, in particular for the higher
luminosity bins (although their noise is larger), an effect that might
be caused by the inclusion of edge-on systems in PSS96 analysis (see
\S~\ref{s_ext}). Beyond \ropt, where the URC points were derived from
linear extrapolations mostly based on \hi\
data, the agreement is good for the 3 lower \Mi\ bins and becomes
increasingly worse at higher luminosities. In particular, the URC
extrapolations for the curves labeled \Mi $< -$22, characterized by
slopes that vary from marginally to strongly {\em declining}, are
inconsistent with our Polyex fits. Interestingly, the results obtained by
PSS96 by co-adding the RCs of their {\em sample B} in 5 luminosity
bins show no indications of a velocity decrease within 2 \ropt. 
In particular, the curve labeled \Mi $= -$22.6 in their Figure B2 
(which would lie between the two uppermost URC tracings in
Figure~\ref{urc}) has an outer gradient $\delta \simeq 0$, indicating
a flat RC slope between 1 and 2 \ropt. As their Figure~3 shows, the
$\delta$ values for the 5 synthetic RCs are systematically larger than
those derived for the individual objects in {\em sample A}; however, they
are also represented with very large error bars and said to be
consistent with the others.
The URC tracings shown in Figure~\ref{urc} (with error bars removed
for clarity) are compared with the \ropt\ template RC data points in
Figure~\ref{urc_2}. The URC solutions qualitatively follow the trend
of the template data in the lower luminosity bins, but do not recover
the {\em shapes} of our average RCs at higher luminosities. For
instance, the data points of the $-$23.8 and $-$23.0 template RCs lie
{\em near} the URC predictions, but the latter clearly do not
represent {\em fits} to the data. In other words, both inner shapes
and outer slopes of those template RCs, as derived from best fits to
the data points, are not reproduced by the URC tracings. The
disagreement is particularly evident for the $-$23.8 template RC,
where a linear fit to its outer region (say, beyond 0.7 \ropt) is
clearly inconsistent with the strongly declining URC slope.

In summary, our template derivation differs from that of PSS96 in
several details, such as the exclusion of the most edge-on systems,
the resampling of the RCs to the same spatial resolution, and the
normalization to a velocity that is both characteristic of the
corresponding luminosity class and computed over a range of radii
rather than at a single point. But most importantly, the larger size
of our data set allows us to probe the velocity fields of the average
RCs beyond 1 \ropt\ without reducing the number of luminosity bins and
with no need for extrapolations. For radial galactocentric distances smaller
than \ropt, where both PSS96 and our derivations are based on a
homogeneous data set (although differently processed), the results are
fairly consistent, when the larger scatter of the URC data and
different selection criteria are taken into account. Beyond \ropt, the
URC extrapolations qualitatively agree with our data only at low luminosities, and
are inconsistent for \Mi $< -$22. 
PSS96 based their extrapolations on a linear fit to the data points in
their Figure~3, where the outer gradient $\delta$ is plotted as a
function of the \bband\ absolute magnitude. While it is unclear why
the outer gradients of the 5 average RCs built from their {\em sample B}
are larger than those of the other points shown (mostly based on \hi\
data), a fit to those 5 points would yield results in better agreement
with our template RCs.

We conclude that late type, high surface brightness spiral galaxies
show no clear indication, {\em on average}, of declining RCs within at
least 4.5 \rd\ (6 \rd\ for most of our \Mi\ bins), even in the
case of the most luminous systems in our sample. 
Of course, this does not contradict claims of declining RCs at larger
distances in the disks.
Most notably, \citet{cvg91} obtained extended \hi\ synthesis RCs 
for two highly-inclined, late spiral galaxies: NGC 2683 and NGC 3521.
Their \iband\ absolute magnitudes, estimated from the \bband\ values 
using a color index $B-I$=1.78 appropriate for their morphological 
type \citep{dej96}, and adopting $\rm H_0=70$ \kmsm, 
are $-$20.6 and $-$22.4, respectively. Both galaxies exhibit declining 
RCs at large radii (the \hi\ extends 15 and 12 optical \rd, 
respectively), and are included by \citet{cvg91} in the category
of ``bright compact galaxies'' based on their small values of \rd.
The two RCs also show no decrease in rotational velocity 
within 6 \rd\ and are thus consistent with our template results.
The maximum rotation velocity of NGC 3521, of the order of 240 \kms,
is consistent with the corresponding template RC;
the fact that NGC 2683, which is $\sim$2 magnitudes 
dimmer, reaches the same maximum velocity is somewhat puzzling but not
wholly surprising, given the expected scatter.
Extended RCs from \hi\ aperture synthesis observations were also
obtained by \citet{spe05} for eight of the fastest known rotators, with
the purpose of investigating the role of disk stability in setting the
observed upper limit to the rotational velocities of spiral galaxies.
This is a very interesting issue, since it is unclear that the most
massive disks can be stabilized by the same processes invoked for less
massive systems; in fact, alternative scenarios have been proposed
(see \citealt{spe05} and references therein), one of which predicts
roughly flat RCs \citep{sm99}. For the eight objects in her sample
(extracted from SFI++), Spekkens produced hybrid \Ha + \hi\ RCs with
extents ranging from 1.8 to 3.3 \ropt. All these RCs are remarkably
flat beyond \ropt\ (about half showing a very mild velocity decline for 
$r> 1.5-2$ \ropt) and in qualitative agreement with our template
solutions.

\section{Discussion and Conclusions}\label{s_concl}

\begin{figure}
\plotone{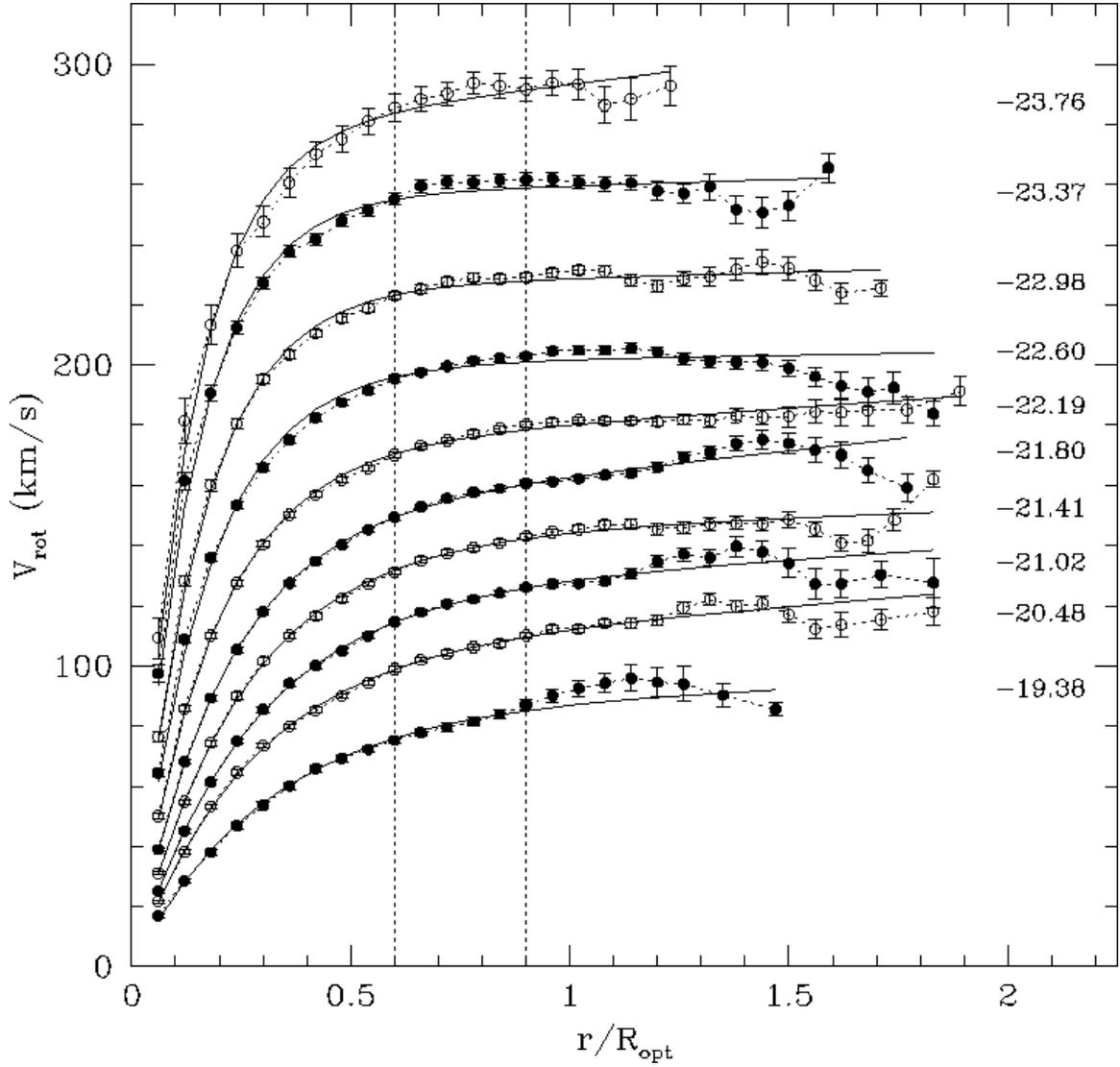}
\caption{Same as Figure~\ref{avgrcs_rd} for template RCs parameterized
  as functions of optical radii. This sample includes 2169 RCs
  extending beyond 0.6 \ropt, and with inclination $i<$80\deg. Polyex
  model coefficients are presented in Figure~\ref{pe_parms_ropt} and
  Table~\ref{mbins_ropt}.
\label{avgrcs_ropt}}
\end{figure}

\begin{figure}
\plotone{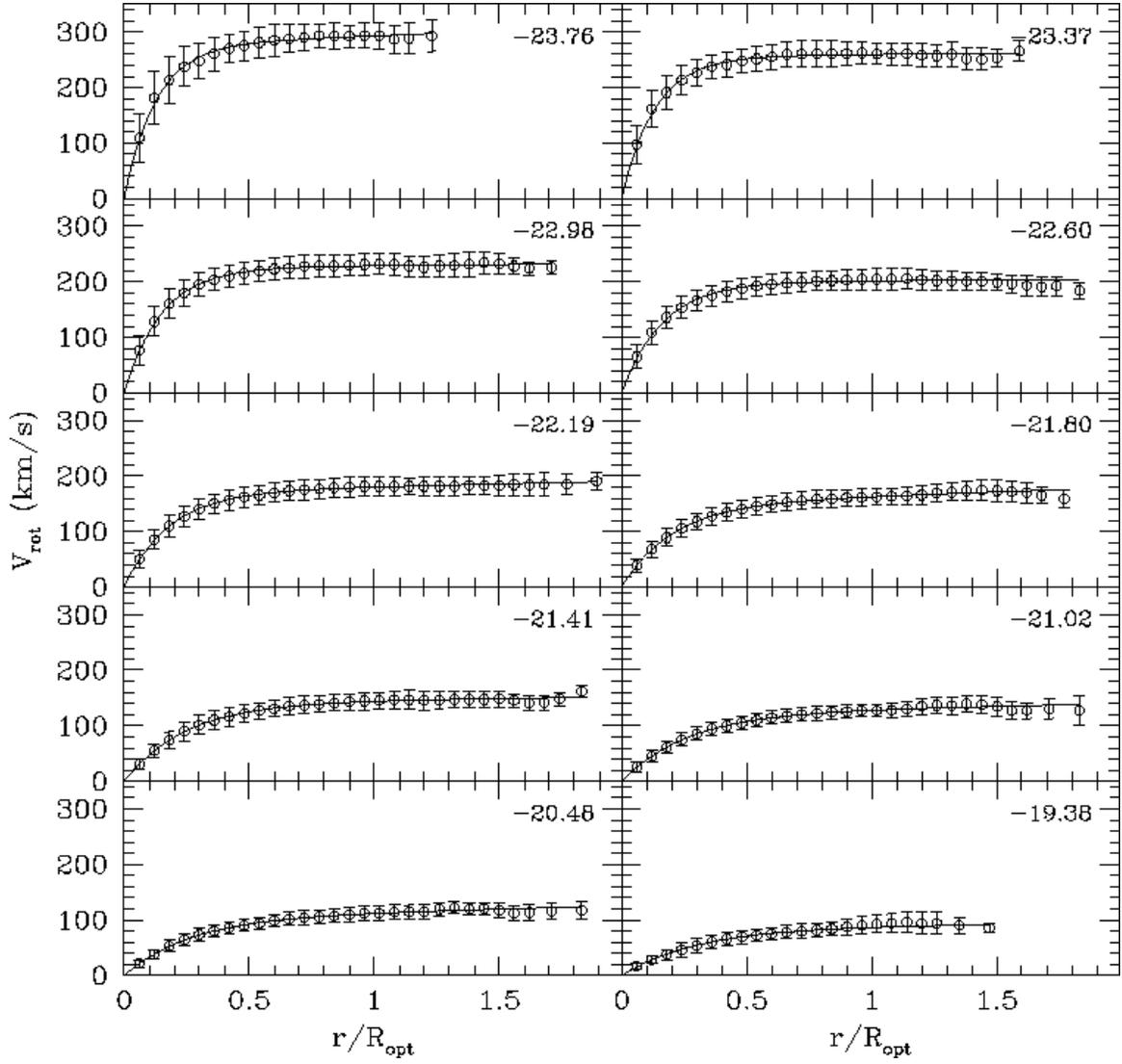}
\caption{Same as Figure~\ref{avgrcs_rd_stdv} for the \ropt\
  parameterization. The solid lines indicate the Polyex fits
  from Figure~\ref{avgrcs_ropt}.
\label{avgrcs_ropt_stdv}}
\end{figure}

\begin{figure}
\plotone{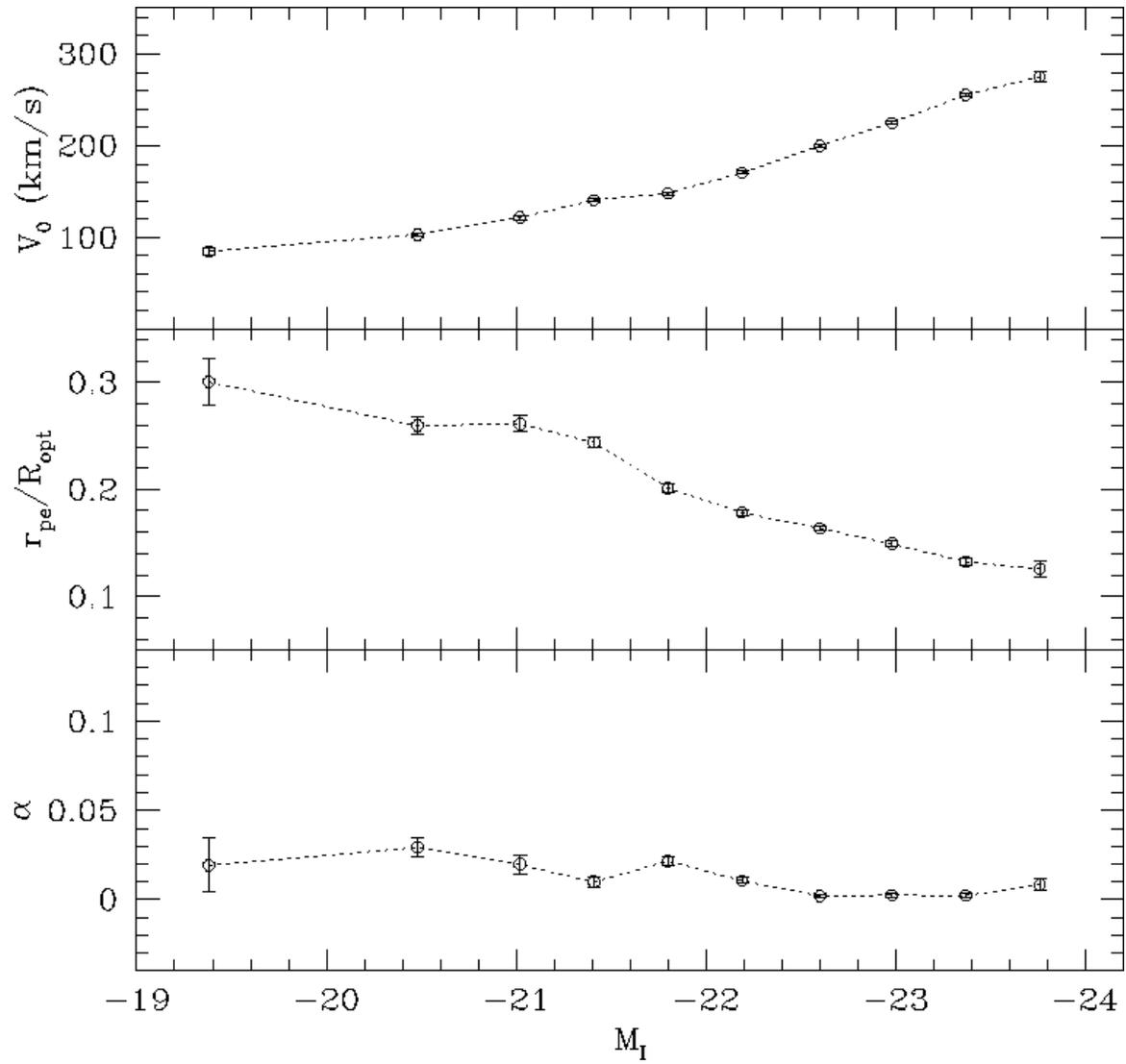}
\caption{Polyex model coefficients of the fits shown in Figure~\ref{avgrcs_ropt}; 
  see also Table \ref{mbins_ropt}. The vertical scales are the same as
  those in Figure~\ref{pe_parms_rd}, except for the central panel.
\label{pe_parms_ropt}}
\end{figure}

\begin{figure}
\plotone{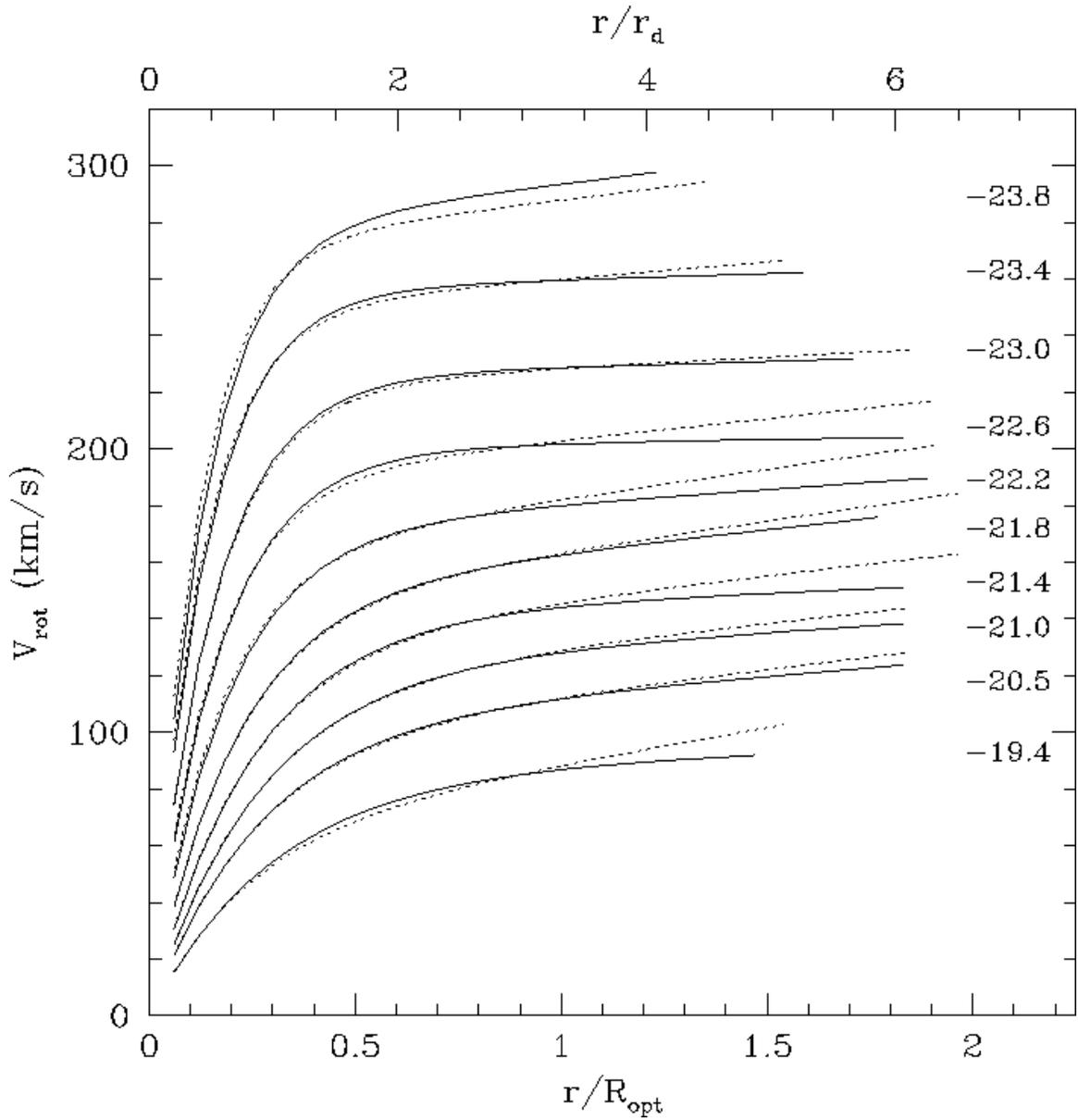}
\caption{Comparison between template RCs expressed in terms of \ropt\
  (solid lines) or \rd\ (dashed). The upper horizontal scale
  corresponds to the lower one multiplied by a factor 3.3, the
  average value of the \ropt/\rd\ ratio for the galaxies of the \rd\ sample.
\label{avgrcs}}
\end{figure}

\begin{figure}
\plotone{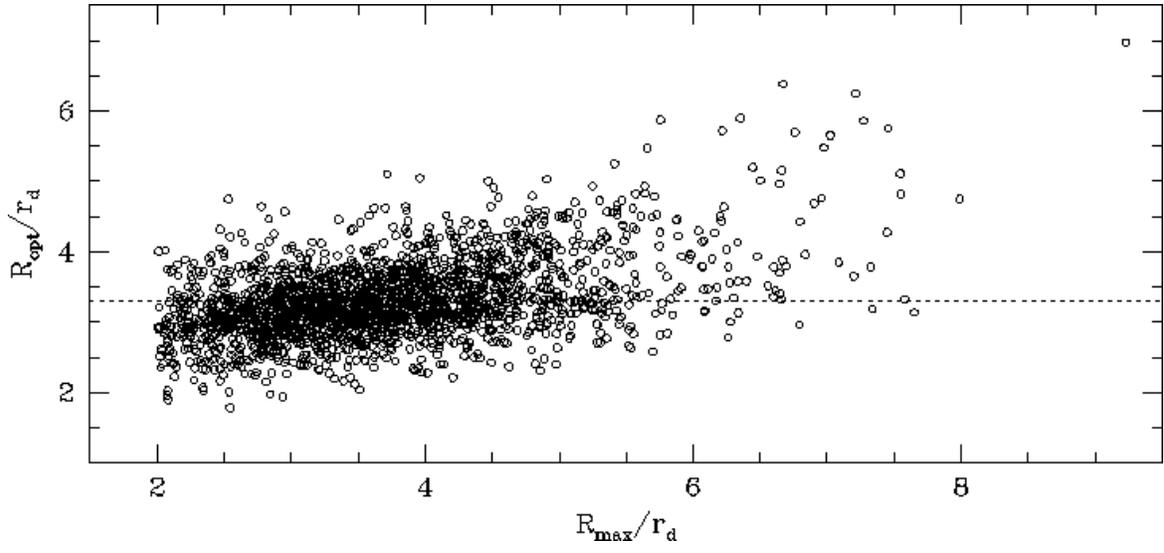}
\caption{Correlation between optical radius and RC extent, both
  expressed in units of \iband\ exponential disk scale lengths, for
  the \rd\ sample. The dotted line is at  \ropt/\rd = 3.3, the
  average value for this data set. The trend observed here accounts for the
  systematic difference of outer slope between the two sets of template RC
  shown in Figure~\ref{avgrcs}.
\label{roptrd}}
\end{figure}

\begin{figure}
\plotone{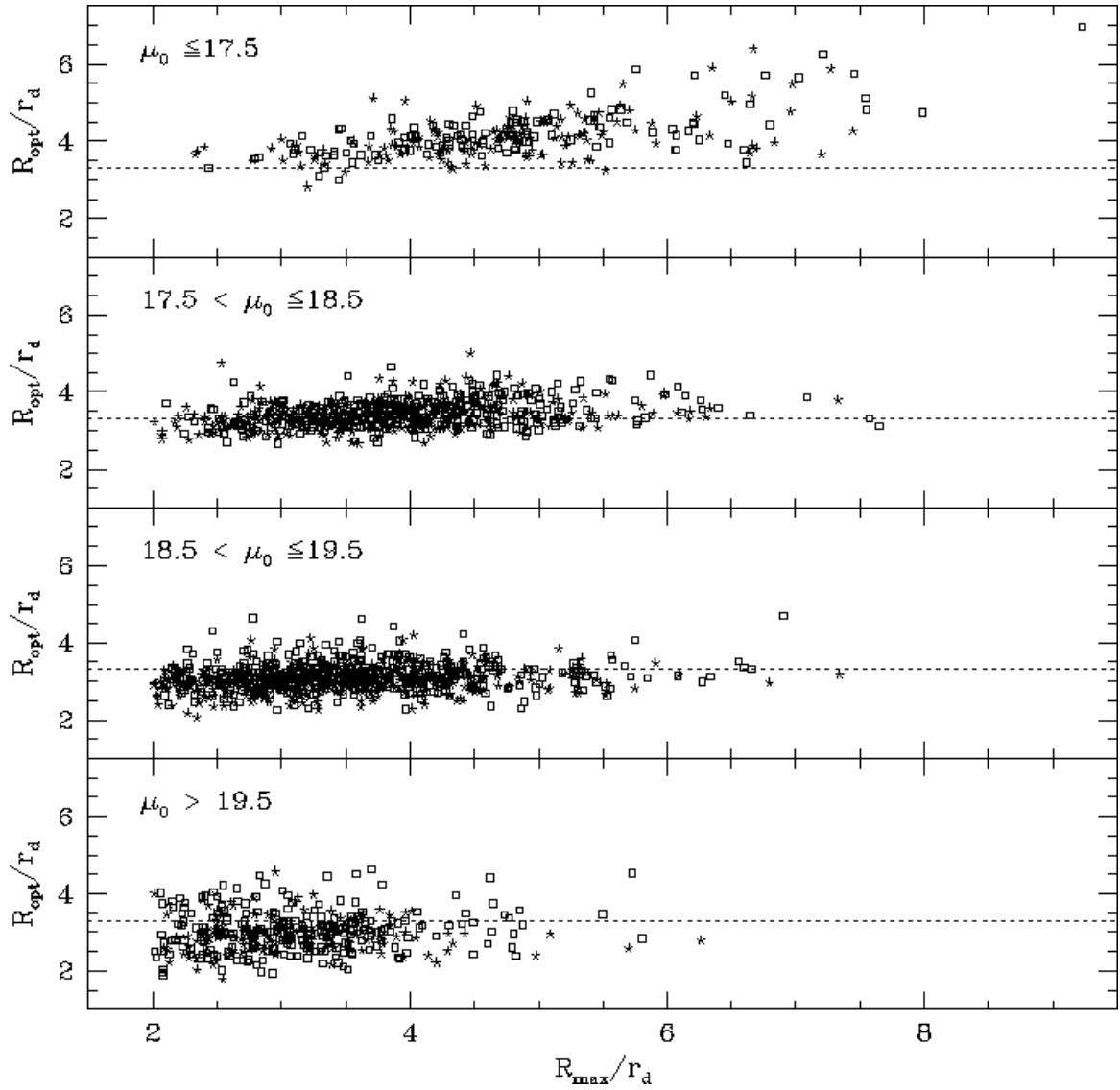}
\caption{Same as Figure~\ref{roptrd}, with the sample divided into
  four bins of central surface brightness, $\mu_0$, as indicated in
  each panel. Galaxies with morphological type Sb or earlier are
  represented by asterisks, those classified as Sbc or later types are
  marked as open squares.
\label{roptrd_sb}}
\end{figure}

\begin{figure}
\plotone{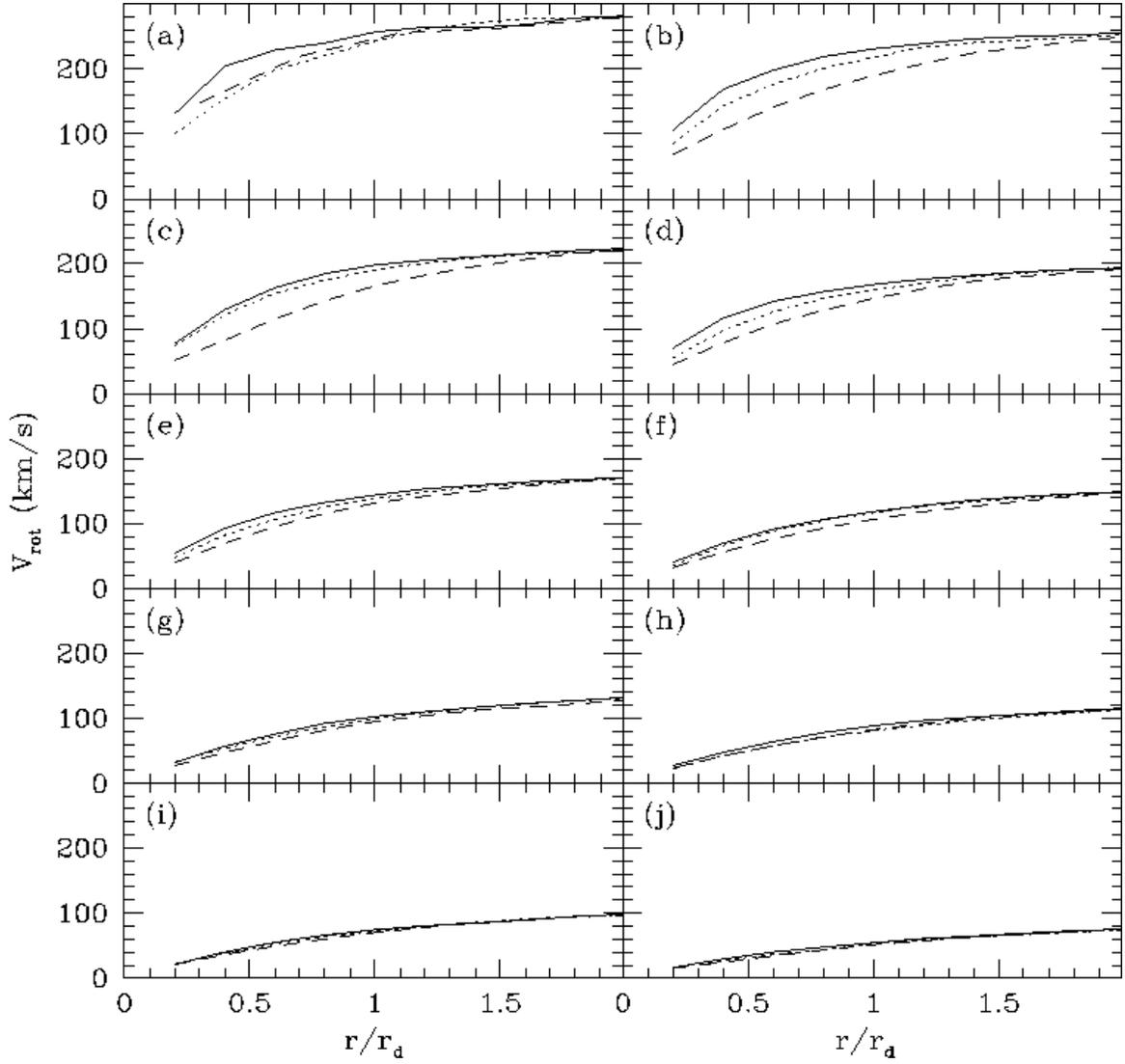}
\caption{Template RCs divided into 3 inclination intervals: 30\deg$\leq i<$70\deg (solid lines), 
  70\deg$\leq i<$80\deg (dotted), and 80\deg$\leq i\leq$90\deg (dashed). 
  Panels (a) through (j) correspond to the luminosity classes \Mi=$-$23.8 to $-$19.0 
  in Table \ref{mbins_rd}. Only the inner regions of the RCs are shown.
\label{incl}}
\end{figure}

Taking advantage of a large sample of long-slit optical RCs with
available \iband\ photometry, we have reinvestigated the dependence of
RC shape on galaxy luminosity.
We determined {\em template} relations by fitting
a function, the Polyex model (eq. \ref{eq_polyex}), to the average RCs
calculated in 10 luminosity classes, spanning the \iband\ absolute
magnitude range [$-$24.0,$-$18.0]. After ascertaining the effect of
internal extinction on our results, we excluded the 472 most edge-on
systems ($\rm i\geq$ 80\deg) from the analysis.
The template RCs are expressed as functions both of exponential disk scale
lengths \rd\ and of optical radii \ropt, the latter to allow a direct 
comparison with PSS96's average RCs and URC predictions.
When scaled with \rd, the template relations show a smooth transition
of RC shapes with increasing \iband\ luminosity, with the most
luminous systems being characterized by steeper inner velocity rises
and flatter outer slopes. The \ropt\ templates show analogous
variations of RC amplitudes and inner slopes with \iband\ luminosity,
but their outer slopes are nearly constant. A direct comparison of the
two sets of templates shows that the results are very similar
within \ropt; beyond that radius the \rd\ templates have steeper
slopes. This difference is mostly attributed to the high brightness
galaxies, whose RCs are typically traced further out in
the disks, and which are characterized by \ropt/\rd\ ratios larger
than the average for the rest of the sample.
We do not find convincing evidence for declining RCs {\em within the
spatial scales sampled by our data}, i.e. 4.5--6 \rd, even for the
most massive systems, in contradiction with the URC models (which,
beyond \ropt, are based on linear extrapolations). 
As argued in \S \ref{s_res}, there are a few examples of
template RCs whose outermost points show a small velocity decrease,
but there is no clear indication of declining outer slopes (as they
would be measured from linear fits to the outer regions of the average
RCs beyond 2 or 3 \rd, or, equivalently, 0.6 or 0.9 \ropt). One
template RC (the one labeled $-$22.60 in Figure \ref{avgrcs_ropt}) does
appear to be declining beyond \about 1.1 \ropt, but the adjacent
curves have rising or flat outer slopes.
For less luminous objects or below \ropt, our templates are qualitatively
consistent with the URC tracings, when the larger scatter of PSS96's
average RCs and different selection criteria (namely, the exclusion of
edge-on galaxies from our analysis) are taken into account.
Being significantly larger than the sample upon which the URC is
based, our data set allows us to compute average RCs that extend
beyond \ropt\ without reducing the number of luminosity bins or
resorting to extrapolations.
Our results are consistent with the declining RCs
of \citet{cvg91}, since the velocity decrease is observed in \hi\ data
that span twice the distance sampled by our optical data, and with the
hybrid \Ha + \hi\ RCs of 8 fast rotators studied by \citet{spe05}, which
show remarkably flat or mildly declining outer slopes beyond 1.5 \ropt.

Based on our results we cannot make quantitative inferences about the
dynamic role of dark matter in galaxies, because optical RCs alone
cannot constrain the properties of dark matter halos on large scales.
In fact, a detailed mapping of the radial distribution of visible and
dark matter in galaxies requires the use of luminosity profiles and
extended \hi\ RCs in addition to optical RCs. 
Nonetheless, the comparison between Figure \ref{avgrcs_ropt} and the URC
models suggests that dark matter plays a more significant role in
determining the internal kinematics {\em within the optical disk} of
the most luminous systems than what claimed by PSS96. At high
luminosity, the profiles of PSS96's RCs show a strong velocity
decrease between 1 and 2 \ropt; their decomposition into a disk and a
dark halo component in terms of the URC parameterization implies a
modest dark matter content, whereas low luminosity systems are dark
matter dominated. Since the templates RCs of the high luminosity
systems do not appear to decline over those spatial scales, their dark
matter content within the optical disk based on the URC predictions
might be underestimated.

The template RCs, expressed by the Polyex parameters as a function of
galaxy luminosity, can be used to constrain models of the circular
velocity field of disks, with applications that range from studies of
dark matter content and kinematic properties of galaxies, to numerical
simulations of disk formation and evolution within the current
cosmological framework, and in general studies that, for different
reasons, must rely on RC models. Our parameterization in terms of disk
scale lengths should prove especially useful in the theoretical and
numerical modeling of disk structure and evolution.

The results presented here will be used to determine statistical
corrections for aperture bias of emission line widths obtained with fiber
spectroscopy, such as those that the on-going Sloan Digital Sky Survey
\citep{str02} is collecting for one million galaxies. Full rotational
velocities of quiescent disk galaxies with apparent sizes larger than the fiber
diameter can be statistically recovered by simulating the impact of
the finite aperture on the observations, where the galaxy RCs are
modeled using our template relations (B. Catinella et al., in
preparation; \citealt{cat05}).

\acknowledgements

\begin{figure}
\plotone{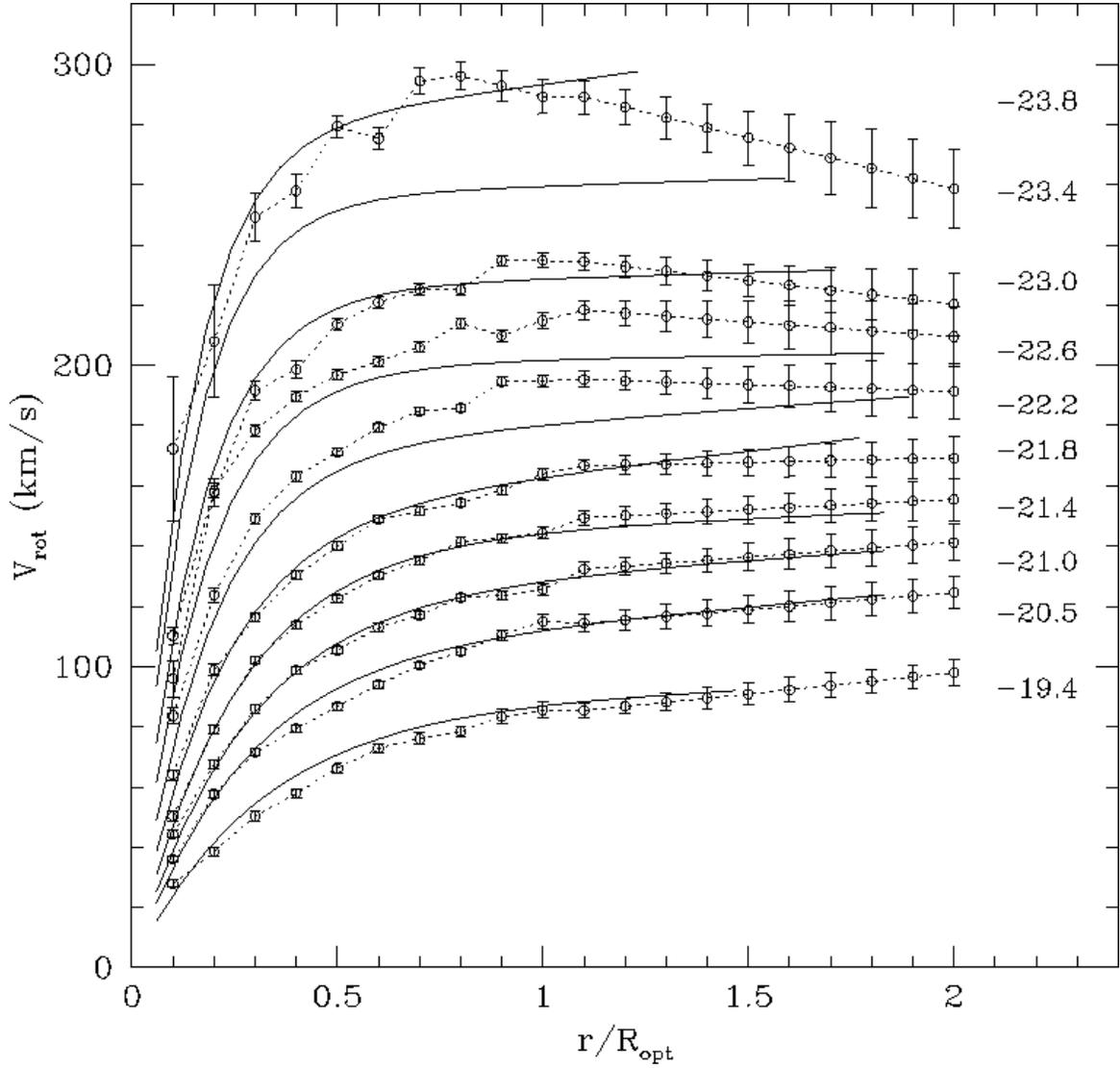}
\caption{Polyex fits (solid lines) to the template RCs shown in
  Figure~\ref{avgrcs_ropt} are here compared to the average RCs
  obtained by PSS96 and kindly provided by Paolo Salucci. All PSS96 data
  points have been shifted by $+$10 \kms\ to match the velocities of
  the template curves at \ropt.
\label{urc}}
\end{figure}

\begin{figure}
\plotone{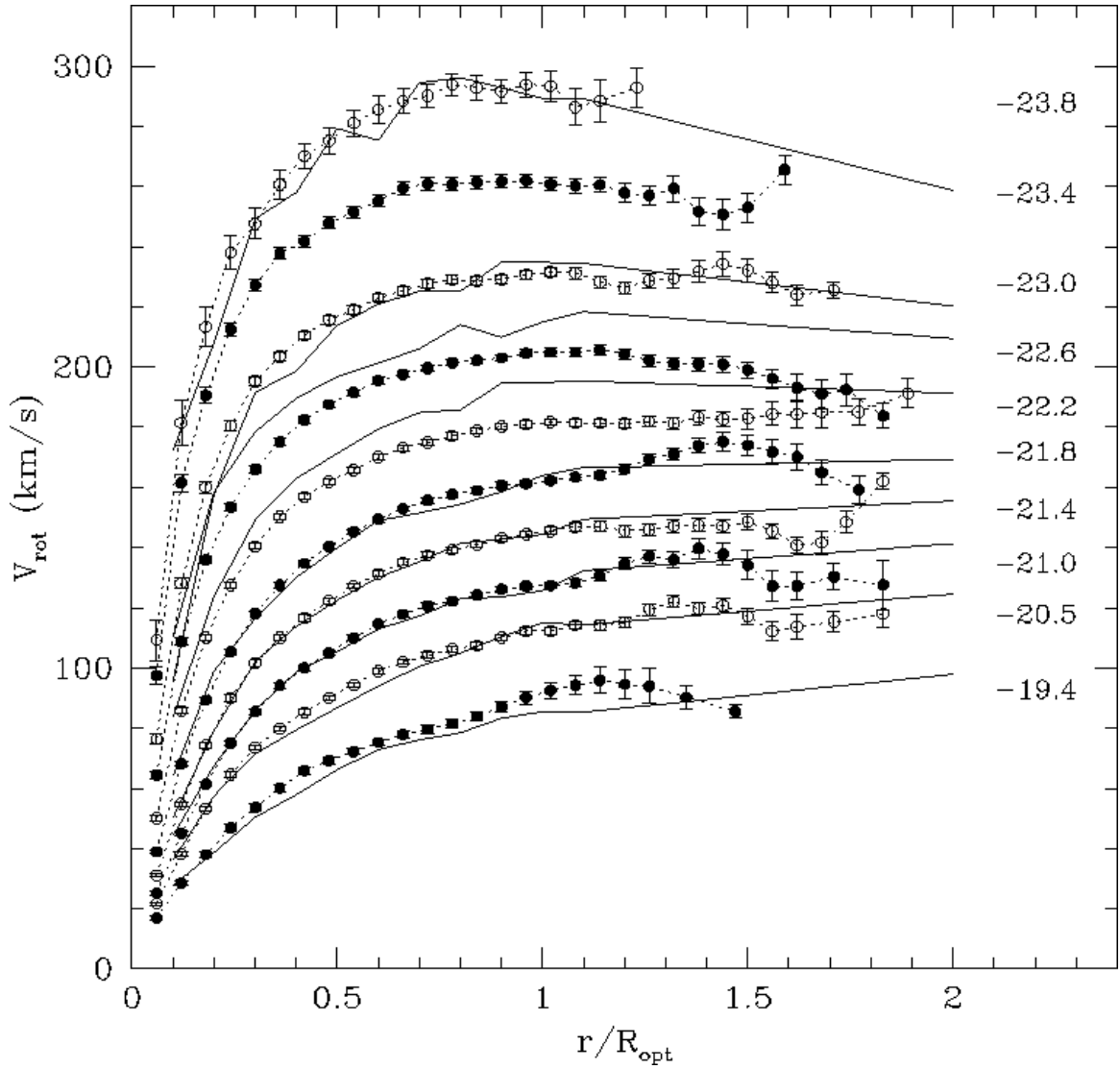}
\caption{The URC tracings in Figure~\ref{urc} are here reproduced as
  solid lines, superimposed on the \ropt\ template RC data from Figure~\ref{avgrcs_ropt}.
\label{urc_2}}
\end{figure}

We thank Paolo Salucci for contributing the data points in Figure~\ref{urc} and
K. L. Masters, K. Spekkens and C. M. Springob for communicating results in advance
of publication. We also thank our anonymous referee for helpful
comments that improved the presentation of this paper.
This research was partly supported by a NAIC pre-doctoral research 
grant at the Arecibo Observatory to BC and by NSF grants AST-9900695
and AST-0307396. \\

\begin{deluxetable}{cccccccc}
\tablecaption{RC Distribution in Luminosity Classes and Polyex Model  Fits to Template RCs 
  Parameterized as Functions of Exponential Disk Scale Lengths\label{mbins_rd}}
\tablewidth{0pt}
\tablehead{
\colhead{\Mi} & \colhead{$\Delta M_{\rm I}$} &  \colhead{$\langle M_{\rm I} \rangle$} & 
\colhead{$\langle v_{\rm \Delta r,M} \rangle$} & \colhead{$N_{\rm \Delta r,M}$} &
\colhead{$V_0$} & \colhead{\rpe / \rd} & \colhead{$\alpha$} \\
\colhead{(1)}&\colhead{(2)}&\colhead{(3)}&\colhead{(4)}&\colhead{(5)} &
\colhead{(6)}&\colhead{(7)}&\colhead{(8)}
}
\startdata
  $-$23.80 &  0.40 &  $-$23.76 &  285.1 & \phn 43 &  270$\pm$5 &  0.37$\pm$0.02 &  0.007$\pm$0.003 \\
  $-$23.40 &  0.40 &  $-$23.37 &  258.4 &  124    &  248$\pm$2 &  0.40$\pm$0.01 &  0.006$\pm$0.001 \\
  $-$23.00 &  0.40 &  $-$22.98 &  225.2 &  225    &  221$\pm$1 &  0.48$\pm$0.01 &  0.005$\pm$0.001 \\
  $-$22.60 &  0.40 &  $-$22.60 &  198.4 &  324    &  188$\pm$1 &  0.48$\pm$0.01 &  0.012$\pm$0.001 \\
  $-$22.20 &  0.40 &  $-$22.19 &  175.7 &  341    &  161$\pm$1 &  0.52$\pm$0.01 &  0.021$\pm$0.001 \\
  $-$21.80 &  0.40 &  $-$21.80 &  155.1 &  327    &  143$\pm$1 &  0.64$\pm$0.01 &  0.028$\pm$0.002 \\
  $-$21.40 &  0.40 &  $-$21.41 &  137.4 &  263    &  131$\pm$1 &  0.73$\pm$0.02 &  0.028$\pm$0.003 \\
  $-$21.00 &  0.40 &  $-$21.02 &  120.7 &  213    &  116$\pm$2 &  0.81$\pm$0.02 &  0.033$\pm$0.005 \\
  $-$20.40 &  0.80 &  $-$20.48 &  103.7 &  203  &\phn 97$\pm$2 &  0.80$\pm$0.02 &  0.042$\pm$0.005 \\
  $-$19.00 &  2.00 &  $-$19.37 &\phn 79.5 &\phn 92 &\phn 64$\pm$3 &  0.72$\pm$0.05 &  0.087$\pm$0.016 \\
\enddata 
\end{deluxetable}

\begin{deluxetable}{cccccccc}
\tablecaption{RC Distribution in Luminosity Classes and Polyex Model Fits to Template RCs 
  Parameterized as Functions of Optical Radii\label{mbins_ropt}}
\tablewidth{0pt}
\tablehead{
\colhead{\Mi} & \colhead{$\Delta M_{\rm I}$} &  \colhead{$\langle M_{\rm I} \rangle$} & 
\colhead{$\langle v_{\rm \Delta r,M} \rangle$} & \colhead{$N_{\rm \Delta r,M}$} &
\colhead{$V_0$} & \colhead{\rpe / \ropt} & \colhead{$\alpha$} \\
\colhead{(1)}&\colhead{(2)}&\colhead{(3)}&\colhead{(4)}&\colhead{(5)} &
\colhead{(6)}&\colhead{(7)}&\colhead{(8)}
}
\startdata
  $-$23.80 &  0.40 &  $-$23.76 &  290.5 & \phn 44 &  275$\pm$6  &  0.126$\pm$0.007 &  0.008$\pm$0.003 \\
  $-$23.40 &  0.40 &  $-$23.37 &  260.6 &  130    &  255$\pm$2  &  0.132$\pm$0.003 &  0.002$\pm$0.001 \\
  $-$23.00 &  0.40 &  $-$22.98 &  227.8 &  226    &  225$\pm$1  &  0.149$\pm$0.003 &  0.003$\pm$0.001 \\
  $-$22.60 &  0.40 &  $-$22.60 &  200.2 &  328    &  200$\pm$1  &  0.164$\pm$0.002 &  0.002$\pm$0.001 \\
  $-$22.20 &  0.40 &  $-$22.19 &  176.3 &  346    &  170$\pm$1  &  0.178$\pm$0.003 &  0.011$\pm$0.001 \\
  $-$21.80 &  0.40 &  $-$21.80 &  156.5 &  330    &  148$\pm$2  &  0.201$\pm$0.004 &  0.022$\pm$0.002 \\
  $-$21.40 &  0.40 &  $-$21.41 &  138.4 &  267    &  141$\pm$2  &  0.244$\pm$0.005 &  0.010$\pm$0.003 \\
  $-$21.00 &  0.40 &  $-$21.02 &  121.4 &  210    &  122$\pm$2  &  0.261$\pm$0.008 &  0.020$\pm$0.005 \\
  $-$20.40 &  0.80 &  $-$20.48 &  105.1 &  195    &  103$\pm$2  &  0.260$\pm$0.008 &  0.029$\pm$0.005 \\
  $-$19.00 &  2.00 &  $-$19.38 & \phn 80.7 & \phn 93 & \phn 85$\pm$5 &  0.301$\pm$0.022 &  0.019$\pm$0.015 \\
\enddata 
\end{deluxetable}

\end{document}